\documentclass[letterpaper, 10 pt, conference]{ieeeconf}  
\IEEEoverridecommandlockouts                              

\usepackage{stfloats}

\usepackage{pgfplots}
\DeclareUnicodeCharacter{2212}{−}
\usepgfplotslibrary{groupplots,dateplot}
\usetikzlibrary{patterns,shapes.arrows}
\pgfplotsset{compat=newest}
\usepackage{tikz}
\usetikzlibrary{shapes, arrows.meta, positioning}

\usepackage{cite}
\usepackage[normalem]{ulem}

\usepackage{times}
\usepackage{url}
\usepackage[hidelinks]{hyperref}
\usepackage[utf8]{inputenc}
\usepackage{graphicx}
\usepackage{amsmath}
\usepackage{amsfonts}
\usepackage{booktabs}
\usepackage{algorithm}
\usepackage{algorithmic}

\usepgfplotslibrary{groupplots}
\usepgfplotslibrary{fillbetween}
\usetikzlibrary{arrows,decorations.pathmorphing,positioning,fit,trees,shapes,shadows,automata,calc} 
\usetikzlibrary{patterns,arrows,arrows.meta,calc,shapes,shadows,decorations.pathmorphing,decorations.pathreplacing,automata,shapes.multipart,positioning,shapes.geometric,fit,circuits,trees,shapes.gates.logic.US,fit, matrix,arrows.meta, quotes}
\usetikzlibrary{backgrounds,scopes}

\usepackage{wrapfig}

\usepackage{mathtools}
\usepackage{accents}

\newtheorem{remark}{Remark}

\newcommand{\rev}[1]{#1}

\title{\LARGE \bf
Autonomous Drifting with 3 Minutes of Data via Learned Tire Models
}

\author{Franck Djeumou$^{1}$, Jonathan Y.M. Goh$^{2}$, Ufuk Topcu$^{1}$, and Avinash Balachandran$^{2}$
\thanks{This material is based on work supported by Toyota Research Institute.}
\thanks{$^{1}$F. Djeumou and U. Topcu are with the University of Texas at Austin, Austin, TX, USA. Email: \texttt{\{fdjeumou, utopcu\}@utexas.edu}}%
\thanks{$^{2}$J. Goh and A. Balachandran are with Toyota Research Institute, Los Altos, CA, USA. Email: \texttt{\{jonathan.goh\}@tri.global}}%
}

\begin{document}

\maketitle

\begin{abstract}
    Near the limits of adhesion, the \rev{forces} generated by a tire \rev{are} nonlinear and intricately coupled. Efficient and accurate modelling in this region could improve safety, especially in emergency situations where high forces are required. To this end, we propose a novel family of tire force models based on neural ordinary differential equations and a neural-\texttt{ExpTanh} parameterization. These \rev{models} are designed to satisfy physically insightful assumptions while also having sufficient fidelity to capture higher-order effects directly from vehicle state measurements. They are used as drop-in replacements for an analytical brush tire model in an existing nonlinear model predictive control framework. Experiments with a customized Toyota Supra show that scarce amounts of driving data -- less than three minutes -- is sufficient to achieve \rev{high-performance} autonomous drifting on various trajectories \rev{with speeds up to 45mph}. Comparisons with the benchmark model show \rev{a $4 \times$ improvement in tracking performance}, smoother \rev{control} inputs, and faster and more consistent computation time.
\end{abstract}
\section{Introduction}\label{sec:intro}
Maximizing tire force usage is critical to safely negotiating highly dynamic situations\rev{, e.g., emergency obstacle avoidance}. Yet, accurately predicting the effective force generated by the four tires on a car is a difficult challenge. Firstly, the tire in isolation has many complex nonlinear phenomenon, including force saturation, camber thrust, and nonlinear load dependence. Indeed, significant effort has gone into developing analytical and empirical models for a \textit{single} tire ~\cite{PACEJKA2012xiii,Svendenius2003ConstructionON, svendenius2007tire, Acosta2017TireLF,Guarneri2008ANM, Xu2022TireFE, Matuko2008NeuralNB}, including the Magic Formula \cite{PACEJKA2012xiii} which is frequently used in industry. Despite its popularity, fitting the many parameters of the Magic Formula is difficult and often requires specialized testing \rev{ and facilities}~\cite{Svendenius2003ConstructionON,svendenius2007tire}.

When attached to a vehicle, the complexity compounds, as every input to these models is coupled into suspension dynamics, weight transfer, and other effects. Many control approaches in the literature thus resort to using a single-track assumption\rev{~\cite{Paden2016ASO,You2017VehicleMA,rajamani2011vehicle, Falcone2007PredictiveAS,Kong2015KinematicAD,Polack2017TheKB}}, where these effects are `lumped' into a single tire model at the front and rear axles, and empirically fit to measured vehicle data. This includes the Fiala brush model~\cite{fiala1954seitenkraften}, which has been experimentally demonstrated in autonomous vehicle control scenarios at the limits of handling, including emergency obstacle avoidance, drifting, and racing ~\cite{Goh2019TowardAV,balachandran2023human,Hindiyeh2011ACF,Subosits2021ImpactsOM,Goh2016SimultaneousSA, goh2019automated}. Although the simplicity aids control development, this single tire lumping often fails to accurately capture the intricate coupling created by higher-order effects.

Neural networks, which have universal approximation properties in the limit, could offer a solution. Black-box and Magic Formula-based neural network models\rev{~\cite{Acosta2017TireLF,Guarneri2008ANM, Xu2022TireFE, Matuko2008NeuralNB, Srinivasan2020EndtoEndVE, Spielberg2019NeuralNV,Williams2017InformationTheoreticMP}} have been explored in the literature. However, they do not retain physics-based guarantees, and none has been tested on a full-size car operating near or at the limits of handling. In general, their complexity has to be balanced against overfitting and computational efficiency, especially when reliable, physically insightful extrapolation is required for real-time control. 

\rev{Our first contribution is to} combine the physical insights of tire models with the modelling power of neural networks. \rev{We propose a novel family} of tire force models based on neural ordinary differential equations (NODE)~\cite{Chen2018NeuralOD,Djeumou2022NeuralNW} and neural-$\texttt{ExpTanh}$, a novel parameterization which uses curves generated by the $\exp (\cdot) \tanh (\cdot)$ function. These are designed to have high fitting fidelity while also incorporating fundamental tire modelling insights \cite{PACEJKA2012xiii,Svendenius2003ConstructionON,svendenius2007tire}, including the friction ellipse constraint and `S-shaped' saturation trend. The NODE model defines a differential equation whose family of solutions includes established models such as the Magic Formula~\cite{PACEJKA2012xiii} and Fiala brush model ~\cite{fiala1954seitenkraften}. Through optimization-based techniques~\cite{Chen2018NeuralOD,Djeumou2022TaylorLagrangeNO}, the model is trained to fit vehicle state measurements. To address the computational complexity of training and evaluating NODE models, which requires integrating a differential equation, we also introduce neural-$\texttt{ExpTanh}$, a subset of the NODE model's solutions. Neural-$\texttt{ExpTanh}$ can be trained efficiently and targets real-time control precisely due to its cheap function and gradient computation time.\looseness=-1

\begin{figure}[t]
\centering
    \includegraphics[width=3.3in]{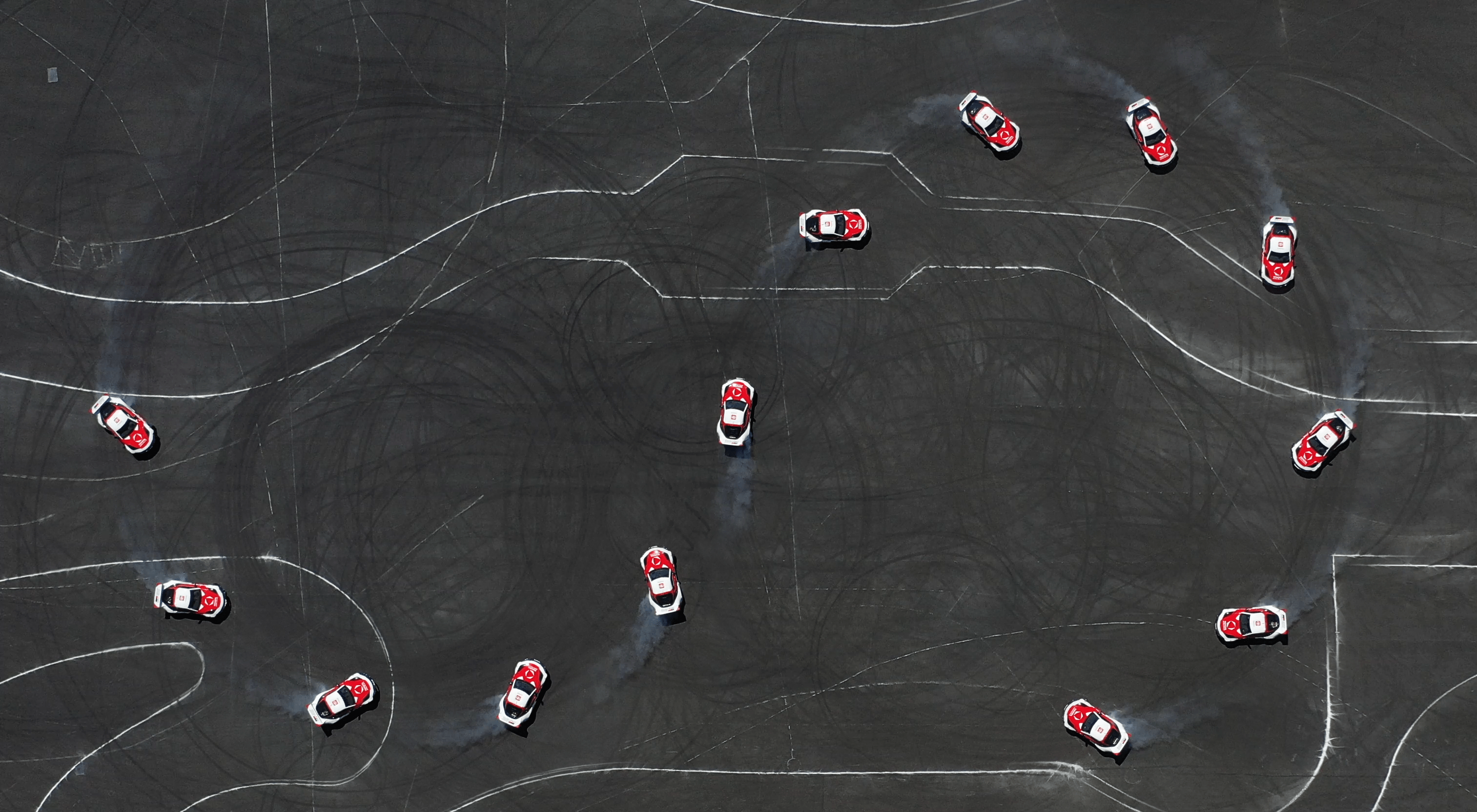}
    \vspace{-2mm}
    \caption{A photocomposite showing stills from an overhead drone video of a fully autonomous experiment superimposed at 1s intervals. The videos of the experiments are available at \textcolor{blue}{\url{https://tinyurl.com/supra-neural-drift}}.}
    \label{fig:supra-fig8}
    \vspace{-6mm}
\end{figure}

\rev{Our second contribution provides an extensive experimental evaluation of these NODE and \texttt{ExpTanh} models on a full-size, heavily-modified Toyota Supra.}
We first compare our models to the Magic Formula and Fiala models on a dataset from the vehicle. The results show that NODE and $\texttt{ExpTanh}$ satisfy the tire fundamentals while being up to $2 \times$ denser than the baselines around zero-mean prediction error. We then use these learned models as drop-in replacements for an analytical brush model in an existing nonlinear model predictive control framework\cite{gohAVEC2022,balachandran2023human}. These are compared in autonomous drifting experiments on two different trajectories which consistently excite the nonlinear regime. Compared to the baseline, the results show improved tracking, fewer steering oscillations, and lower computation time. 

The last set of experiments demonstrates data efficiency and generalization of \rev{our models}. We switch to a different set of tires, collect 3 minutes of manual driving data, train an \texttt{ExpTanh} model in a few seconds, and then perform figure-8 autonomous drifting experiments, \rev{shown in Figure~\ref{fig:supra-fig8}}. The learned model shows similarly good closed-loop performance, while the performance of the \rev{baseline} model drops.
\section{Fundamentals}\label{sec:preliminaries}
\begin{wrapfigure}{r}{1.7in}
    \vspace*{-3mm}
	\includegraphics[width=1.7in, height=1.2in]{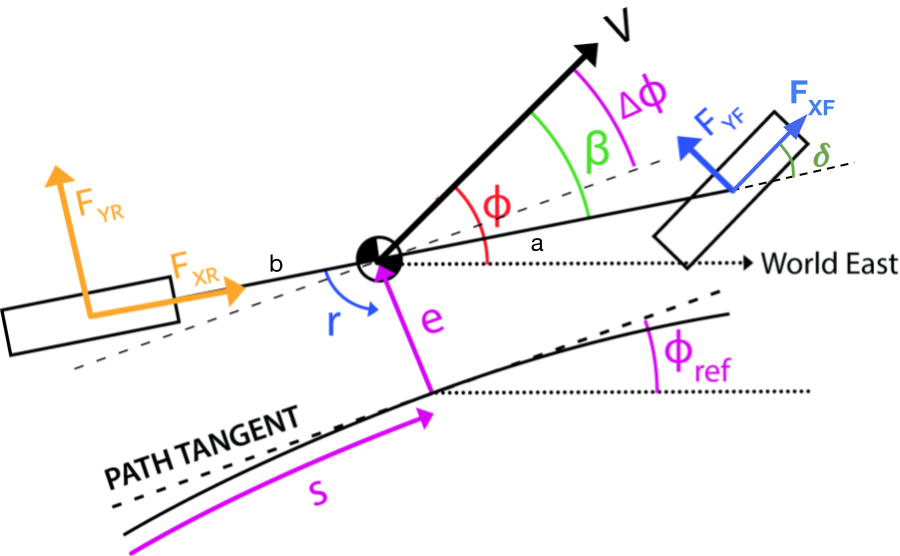}
	\vspace*{-7mm}
	\caption{Single-track model of a vehicle on a reference path.}
	\label{fig:fundamentals}
	\vspace*{-5mm}
\end{wrapfigure}

\textbf{Vehicle Dynamics.}
The dynamics are described using a planar single-track model~\cite{Paden2016ASO,rajamani2011vehicle, Falcone2007PredictiveAS,Kong2015KinematicAD,Polack2017TheKB}, shown in Figure~\ref{fig:fundamentals}. 
They are expressed in a curvilinear coordinate system~\cite{Goh2019TowardAV,Goh2016SimultaneousSA,Subosits2021ImpactsOM}, where the vehicle position is relative to a reference trajectory. The position coordinate is described by the distance $s$ along the path, the relative heading $\Delta \phi$ with respect to a planned course $\phi_{\mathrm{ref}}$, and the lateral deviation $e$ from the path. The motion of the state $x= [r, V, \beta, \omega_{f,r},s, e, \Delta \phi]$ is given by $\dot{x} = M(x,u) [F_{xf}, F_{yf}, F_{xr}, F_{yr}]$, where the matrix $M$ can be found in~\cite{Goh2019TowardAV}. The states $r, V, \beta$, and $\omega_{f,r}$ denote, respectively, the yaw rate, velocity, sideslip angle, and front/rear axle wheelspeed. The components of the control $u = [\delta, \tau_{f,r}]$ represent the steering angle and the torque exerted on the front/rear axle. Further, $F_{xf}, F_{yf}, F_{xr}$, and $F_{yr}$ define, respectively, the front longitudinal, front lateral, rear longitudinal, and rear lateral tire forces. In what follows, we use $F_x$ or $F_y$ to refer to $F_{xf}, F_{xr}$ or $F_{yf}, F_{yr}$, where it is clear from the context.

\textbf{Tire Force Fundamentals.}
Modeling the nonlinear tire forces $F_x, F_y$ has been explored extensively. Many models consider these forces to be generated by relative `slip' velocity between the \rev{tire} contact patch and the road. The slip angle $\alpha$, longitudinal slip ratio $\sigma$, and total slip $\kappa$ as given by~\eqref{eq:slip-angles} are often used as model inputs \rev{for the tire forces}:
\begin{align}
    &\alpha_f = \arctan \frac{ V\sin \beta + a r}{V \cos \beta} - \delta, \; \alpha_r =  \arctan \frac{ V\sin \beta - b r}{V \cos \beta} \nonumber \\
    &\sigma_{f,r} = \frac{r_w \omega_{f,r} -  V_{xf,xr}}{V_{xf,xr}}, \;\; \kappa_{f,r} = \sqrt{\tan \alpha_{f,r}^2 + \sigma_{f,r}^2}
    \label{eq:slip-angles}
\end{align}
where $V_{xf} = V \cos (\delta - \beta) -  a r \sin{\delta}$, $V_{xr} =  V \cos \beta$, \rev{$r_w$ is the effective radius of the wheel}, and $a,b$ are the distances from the center of gravity to the front and rear axles. 
We follow the sign convention of $F_y \leq 0$ for nonnegative $\alpha$ and $F_x \geq 0$ for nonnegative $\sigma$.

\textbf{Measurements of Tire Forces.}
In this work, we learn tire models from the state measurements and estimates of the effective lumped axle forces. While there are different strategies for estimating these forces, for simplicity in this paper, we consider conditions where we can assume $F_{xf} = 0$, e.g., no torque on the front wheels. 
Then, we \rev{compute} $\dot{r},\dot{V},\dot{\beta}$ from measured states and invert through the matrix $M$ \rev{to obtain estimated forces, indicated by $\Bar{F}$ ~\cite{Goh2019TowardAV}}.
\section{Physics-Informed Learned Tire Forces}\label{sec;physics-model}
\begin{figure}[t]
    \centering
    \includegraphics[width=0.9\linewidth, ]{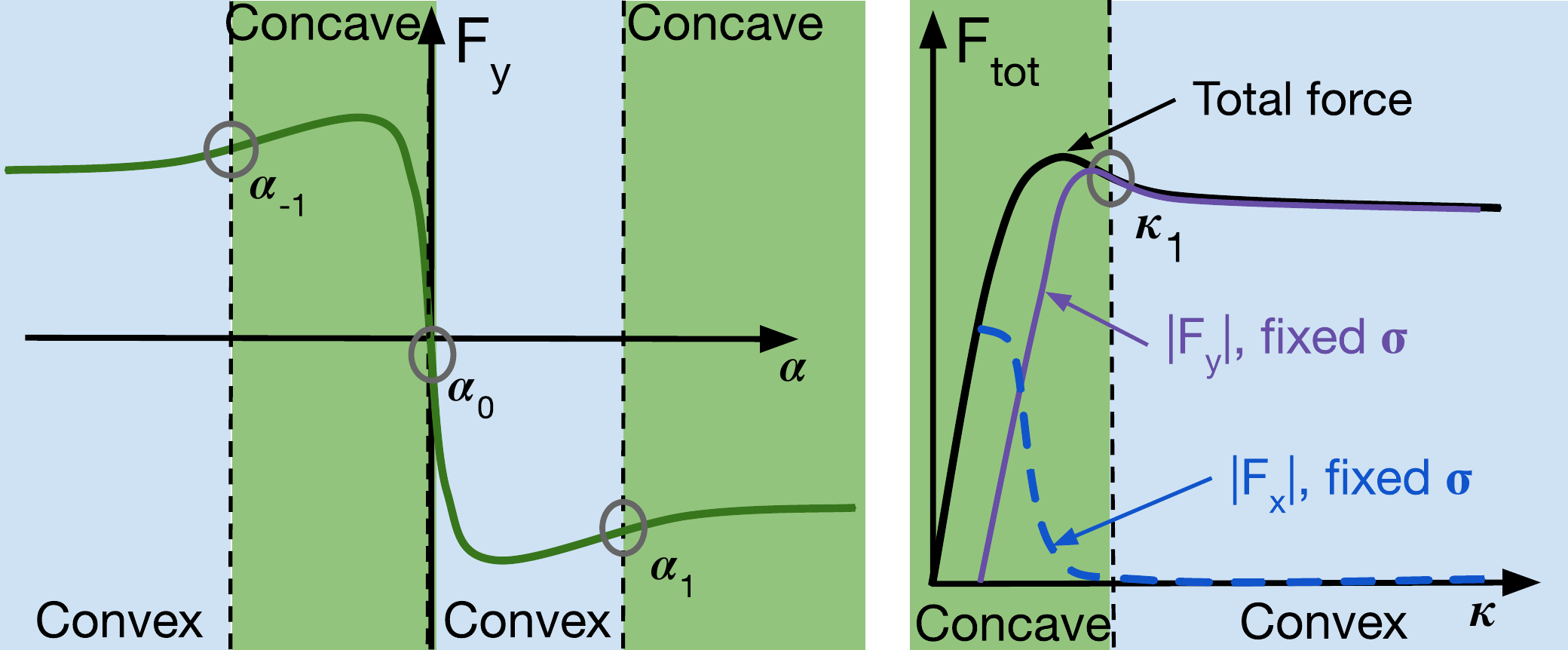}
    \vspace*{-2mm}
    \caption{The left figure shows the inflection points $\alpha_{-1}, \alpha_0, \alpha_{1}$ and the changes in the convexity/concavity of $F_y$ in the pure slip regime. The right figure shows the inflection point $\kappa_1$ of $F_{\mathrm{tot}}$ in the combined slip regime.}
    \vspace*{-5mm}
    \label{fig:inflexion}
\end{figure}

In this section, we describe our physics-based, neural ordinary differential equation (NODE) model and the derived \texttt{ExpTanh} parameterization. 
\rev{In keeping with tire modelling convention, we divide the discussion into pure slip and combined slip regimes. In pure slip, the tire is only creating force along one axis ($\sigma=0$ or $\alpha=0$); in combined slip, the tire is creating both longitudinal and lateral force ($\sigma\neq0$ and $\alpha\neq 0$).}
From tire fundamentals~\cite{PACEJKA2012xiii,Svendenius2003ConstructionON,svendenius2007tire}, we expect the following generalized behavior, summarized in Figure~\ref{fig:inflexion}.

\textbf{\rev{Characteristic} `S-shape' curve.}
As the absolute value of the input slip increases, the tire force magnitude also increases until a peak force is attained and the \rev{tire} contact patch starts to slide. Beyond this point, \rev{the} force decreases, following an `S-shape' curve. In the pure slip \rev{regime}, the input slip is $\sigma$ or $\alpha$. In the combined slip \rev{regime}, the input slip is some combination of $\sigma$ and $\alpha$, and the output is the total force magnitude $F_{\mathrm{tot}} = \sqrt{F_x^2 + F_y^2}$.

\textbf{Combined slip regime.}
For combined slip, the components of $F_{\mathrm{tot}}$ are distributed according to some ratio of the slip angle and longitudinal slip vs. the combined slip. An example is schematically shown in Figure~\ref{fig:inflexion}, for fixed $\sigma$ and $|\alpha|$ increasing from 0: The proportion of longitudinal force decreases while the lateral force increases until saturation. 

\textbf{Friction \rev{limits}.}
In both regimes, the peak force is constrained by the maximum available tire/road adhesion capability, $\mu F_z$, with $\mu$ the friction coefficient and $F_z$ the normal load on the tires. $\mu F_z$ is difficult to know precisely as it depends on the surface, tire orientation, and the normal load -- which in turn vary with the vehicle's state due to weight transfer/suspension dynamics.
Yet, this notion of a maximum force greatly eases analysis for control and safety.

Throughout this paper, we assume a given set of measurements $\mathcal{D} = \{(\alpha_{f,r},\sigma_{f,r},r,V,\beta,\omega_{f.r}, \Bar{F_x}, \Bar{F_y}, \Bar{\mu F_z})_i\}_{i=1}^N$, \rev{where $\Bar{\mu F_z} = \Bar{\mu} m g$ is a rough estimate of the nominal load and $\Bar{\mu}$ encodes any available approximate knowledge on $\mu$.}

\subsection{Physics-Informed NODE for Tire Force Modeling}\label{subsec:node}
We seek a generalizable model that satisfies these physical insights. First, for the characteristic `S-shape', instead of intuiting a curve, e.g., Magic Formula, that satisfies the `S-shape', we characterize the family of physically-feasible curves using notions of convexity, concavity, and inflection points. Then, we optimize for the function in this family that best fits the data via stochastic gradient descent.

Specifically, at critical inflection points (Figure~\ref{fig:inflexion}), the curve changes convexity or concavity. In the pure slip regime, $F_y$ contains \emph{three} inflection points $\alpha_{-1}, \alpha_0, \alpha_1$.
\rev{We seek a family of curves such that $F_y(\alpha)$ is convex for all $\alpha \leq \alpha_{-1}$ and $\alpha \in [\alpha_{0}, \alpha_1]$, and $F_y(\alpha)$ is concave otherwise, for some $\alpha_{-1}, \alpha_0, \alpha_{1}$.
}
\rev{We have the same properties for $-F_x(\sigma)$, for some $\sigma_{-1}, \sigma_0, \sigma_{1}$.}
In the combined slip regime, the family of curves for $F_{\mathrm{tot}}$ contains an inflection point $\kappa_1$ such that $\forall \kappa \leq \kappa_1$, $F_{\mathrm{tot}}(\kappa)$ is concave, and convex otherwise. 
Convexity and concavity correspond to nonnegative and nonpositive second-order derivatives, respectively. Thus, the main idea is to learn the inflection points and the second derivative of the tire forces with respect to the corresponding slips while enforcing the desired convexity/concavity properties; the forces are then obtained by integration. Further, we enforce soft constraints on the peak force as required by the friction limits.\looseness=-1

\textbf{Pure Slip NODE Model.} 
The lateral force $F_y$ is a solution of the second-order differential equation given by
\begin{align}
    \dot{F}^\theta_y &= G^\theta_{y}, \; z = [\alpha, F^\theta_y, \rev{G^\theta_{y}}, \alpha^\theta_{-1},\alpha^\theta_0, \alpha^\theta_1]\label{eq:fy-node}\\
    \dot{G}^\theta_{y} &= 
    \begin{cases} 
        \mathrm{exp}\{\mathrm{NN}_1^\theta(z, \mathrm{feat}) \} \text{ if } \alpha \leq \alpha^\theta_{-1} \text{ or } \alpha \in [\alpha^\theta_0, \alpha^\theta_1]\\
        - \mathrm{exp}\{\mathrm{NN}_2^\theta(z, \mathrm{feat}) \} \text{ otherwise}
    \end{cases}\nonumber
\end{align}
where the derivative here is taken with respect to $\alpha$. The set of features used for learning are $\mathrm{feat} = [r,V,\beta, \Bar{\mu F_z}]$ for the front axle and $\mathrm{feat} = [r,V,\Bar{\mu F_z}]$ for the rear. 
\rev{We select the feature set $\mathrm{feat}$ such that for fixed $\mathrm{feat}$, $\alpha$ given in~\eqref{eq:slip-angles} is not uniquely defined.}
$\mathrm{NN}_x^\theta$ denotes a neural network, where $\theta$ is the set of all parameters for the model. The inflection points are parameterized as $[\alpha^\theta_{-1}, \alpha^\theta_0, \alpha^\theta_1, F^\theta_0, G^\theta_{0}] = \mathrm{NN}^\theta_3(\mathrm{feat})$ with $F^\theta_0$, $G^\theta_{0}$ being the initial states to use when integrating the differential equation. Note that choosing the $\mathrm{exp}$ function in  $\dot{G}^\theta_{y}$ enforces the nonnegative and nonpositive second-order derivatives constraints. 
We then compute the parameters $\theta$ by solving the following optimization problem
\begin{align}
    \underset{\theta}{\mathrm{min}} \: \frac{1}{N} \sum_{\substack{\alpha,r,V,\beta,\\ \Bar{F_y},\Bar{\mu F_z}  \in \mathcal{D}} } & \Big(\underbrace{\mathrm{ode}\big(\eqref{eq:fy-node}, [\alpha^\theta_0, \alpha], [F^\theta_0, G^\theta_{0}] \big)}_{[F_y^\theta,G_{y}^\theta ]}- \Bar{F_y} \Big)^2 \nonumber \\ 
    &+ \lambda (\min \{\Bar{\mu F_z}- |F^\theta_y|, 0 \})^2 \label{eq:pure-node-opt}
\end{align}
where $\mathrm{ode}$, an integration scheme, solves~\eqref{eq:fy-node} from $\alpha^\theta_0$ to the measured $\alpha$ with the initial condition given by $[F^\theta_0, G^\theta_{0}]$. The term  $\lambda (\min \{\Bar{\mu F_z}- |F^\theta_y|, 0 \})^2$ enforces the friction limits knowledge as a soft constraint by penalizing values that exceed the estimated nominal load $\Bar{\mu F_z}$. The hyperparameter $\lambda$ specifies the confidence in $\Bar{\mu F_z}$: Low values enable the peak force to be adjusted according to the data while high values constrain the peak force to be less than $\Bar{\mu F_z}$.

\textbf{Combined Slip NODE Model.}
The total force $F_{\mathrm{tot}}$ is a solution of the second-order differential equation given by
\begin{align}
    \dot{F}^\theta_{\mathrm{tot}} &= G^\theta_{\mathrm{tot}}, \; z = [\kappa, F^\theta_{\mathrm{tot}}, G^\theta_{\mathrm{tot}}, \kappa^\theta_1]\nonumber\\
    \dot{G}^\theta_{\mathrm{tot}} &= 
    \begin{cases} 
        - \mathrm{exp}\{\mathrm{NN}_1^\theta(z, \mathrm{feat}) \} \text{ if } \kappa \leq \kappa^\theta_{1} \\
        \mathrm{exp}\{\mathrm{NN}_2^\theta(z, \mathrm{feat}) \} \text{ otherwise}
    \end{cases}\label{eq:ftot-node}
\end{align}
where the derivative is with respect to the combined slip $\kappa$, $[\kappa^\theta_{1}, F^\theta_0, G^\theta_{0}] = \mathrm{NN}^\theta_3(\mathrm{feat})$, and the features $\mathrm{feat}$ are again picked such that $\alpha$ and $\sigma$ are not uniquely defined. \rev{Then, to learn the component distribution of this total force,} we define $[s^\theta_1, s^\theta_2] = \mathrm{NN}_4^\theta(\alpha, \sigma)$ and estimate $F^\theta_y$ and $F^\theta_x$ by scaling $F^\theta_{\mathrm{tot}}$ as follows
\begin{align}
    F^\theta_y = \frac{s^\theta_1 F^\theta_{\mathrm{tot}}}{\sqrt{(s^\theta_1)^2 + (s^\theta_2)^2}}, \; F^\theta_x = \frac{s^\theta_2 F^\theta_{\mathrm{tot}}}{\sqrt{(s^\theta_1)^2 + (s^\theta_2)^2}} \label{eq:fxfy-combined}
\end{align}
Thus, the parameters $\theta$ are obtained by solving the problem
\begin{align*}
    &\underset{\theta}{\mathrm{min}} \: \frac{1}{N} \sum_{\substack{\alpha,\sigma,r,V,\beta,\Bar{F_y},\\\Bar{F_x},\Bar{\mu F_z}  \in \mathcal{D}} }  \Big(\underbrace{\mathrm{ode}\big(\eqref{eq:ftot-node}, [\kappa^\theta_0, \kappa], [F^\theta_0, G^\theta_{0}] \big)}_{[F_\mathrm{tot}^\theta,G_{\mathrm{tot}}^\theta ]}- \Bar{F}_{\mathrm{tot}} \Big)^2 \nonumber \\ 
    &+ (F^\theta_y - \Bar{F_y})^2 + (F^\theta_x - \Bar{F_x})^2 +
    \lambda (\min \{\Bar{\mu F_z}- F^\theta_\mathrm{tot}, 0 \})^2 
\end{align*}
where the measured total force is $\Bar{F}_\mathrm{tot}  = \sqrt{\Bar{F_x}^2 + \Bar{F_y}^2}$.

\begin{figure*}[!b]
    \vspace*{-4mm}
    \centering
    \includegraphics{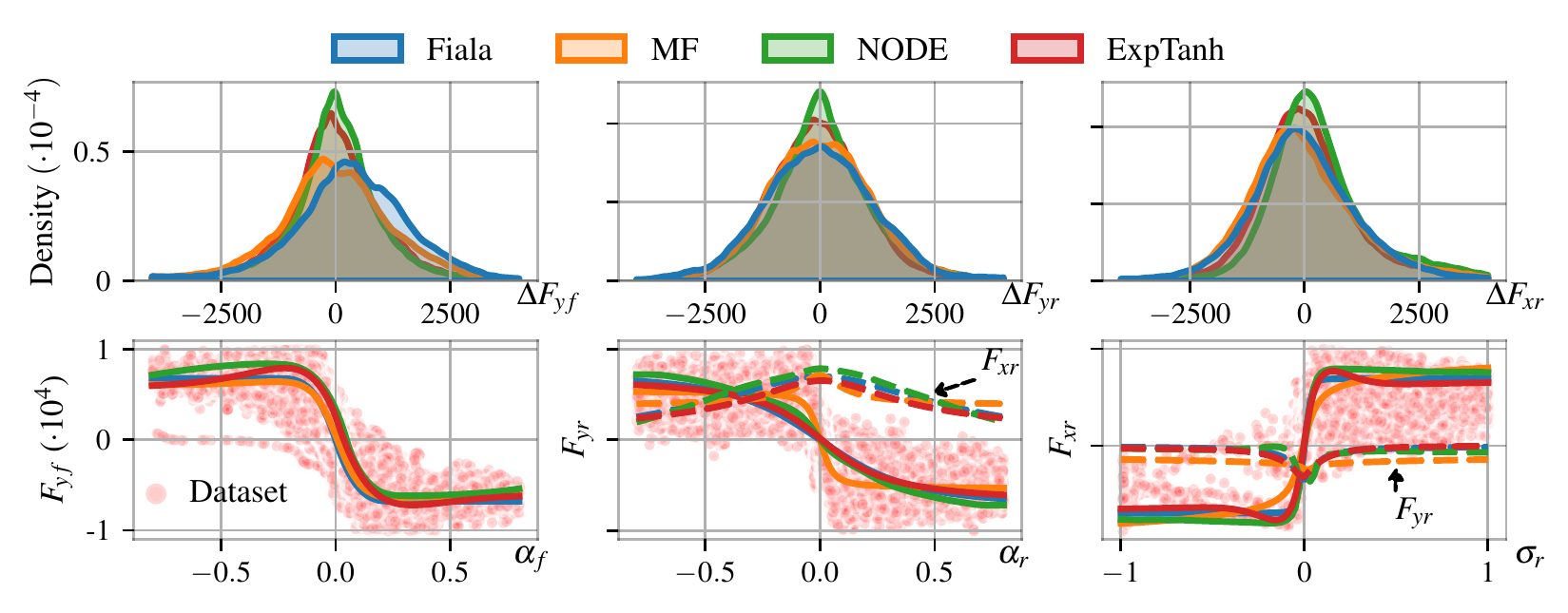}
    \vspace*{-4mm}
    \caption{Comparison of the different tire models trained and tested on a real-world driving dataset. The first row shows the density distribution of the prediction error, and the second row shows the forces as a function of the slip values for a fixed state $r,V,\beta=0.7, 20, 0.1$, where $F_{yr}(\alpha_r)$ is obtained for fixed $\sigma_r = 0.4$ and $F_{xr}(\sigma_r)$ is obtained for fixed $\alpha_r = 0.02$. In the density plot, NODE and $\texttt{ExpTanh}$ are at least $1.5 \times$ denser around zero-mean error than expert-designed Fiala and Magic Formula (MF). The second row validates that the learned NODE and $\texttt{ExpTanh}$ enforce the tire fundamentals.}
    \label{fig:compare-learned-forces}
\end{figure*} 

\begin{remark}\label{rem:node-compute}
    In the pure slip regime, learning $-F^\theta_x$ follows exactly the description of $F^\theta_y$ with $\alpha$ replaced by $\sigma$. Despite the rich class of functions encoded by the NODE formulation, solving~\eqref{eq:fy-node} and~\eqref{eq:ftot-node} to estimate the forces slows down training and hinders the direct application of the formulation for control. In practice, we address this issue by first learning the parameters $\theta$, then training a new neural network to mimic the solutions of~\eqref{eq:fy-node} and~\eqref{eq:ftot-node} via overfitting. 
    Thus, evaluating the obtained neural network and its Jacobian becomes computationally cheap for real-time control.
\end{remark}
\subsection{\texttt{ExpTanh}: A New Family of Tire Models}\label{subsec:exptanh}
We restrict the NODE model's set of solutions to a family of functions, namely \texttt{ExpTanh}, satisfying the second-order derivative condition without the need to integrate a differential equation. \texttt{ExpTanh} curves are given by
\begin{align*}
    \texttt{ExpTanh}^\theta(z) = a^\theta_1 +  a^\theta_2e^{-a^\theta_3|z|} \tanh \big( a^\theta_4(z-a^\theta_5) \big) 
\end{align*}
where $a^\theta_k$ are constants or neural network functions such that $a^\theta_3, a^\theta_4 > 0$. Importantly, the maximum/minimum values $z^\theta_{+}, z^\theta_-$ can be found analytically:
\begin{align}
    z^\theta_{\pm} = a^\theta_5 \pm \frac{1}{a^\theta_4} \mathrm{atanh}\big( \frac{\sqrt{(a^\theta_3)^2 + 4(a^\theta_4)^2} - a^\theta_3}{2a^\theta_4} \big)\label{eq:exptanh-amaxmin}
\end{align}

\textbf{\texttt{ExpTanh} Pure Slip.}
We model $F_y$ as $F^\theta_y(\alpha, \mathrm{feat}) = \texttt{ExpTanh}^\theta(\alpha)$, where \rev{$(a^\theta_i)_{i=1}^5 = \mathrm{NN}^\theta(\mathrm{feat})$}, $\mathrm{feat}$ are the same features as in the NODE version, and $\theta$ is the set of all parameters. In practice, we pass $a^\theta_3, a^\theta_4$ through an exponential function to enforce nonnegative values. The optimum parameters $\mathrm{\theta}$ are given by 
\begin{align*}
    \underset{\theta}{\mathrm{min}} \: \frac{1}{N} \sum_{\substack{\alpha,r,V,\beta,\\ \Bar{F_y},\Bar{\mu F_z}  \in \mathcal{D}}} \Big( F^\theta_y - \Bar{F_y} \Big)^2  + \lambda (\Bar{\mu F_z} - |F_y^\theta(\alpha^\theta_\pm, \mathrm{feat})|)^2
\end{align*}
where the second term is a similar soft penalty on exceeding the estimated maximum friction force.

\textbf{$\texttt{ExpTanh}$ Combined Slip.}
We model the total force as $F^\theta_{\mathrm{tot}}(\kappa, \mathrm{feat}) = \texttt{ExpTanh}^\theta(\kappa)$, where \rev{$(a^\theta_i)_{i=1}^5 = \mathrm{NN}^\theta(\mathrm{feat})$} and $\mathrm{feat}$ represents the same set of features as in the combined slip NODE model. The forces $F^\theta_y$ and $F^\theta_x$ depend on $F^\theta_{\mathrm{tot}}$ as given by~\eqref{eq:fxfy-combined}, where the functions $s_1^\theta$ and $s_2^\theta$ are to be learned. Specifically, we compute $\theta$ by solving 
\begin{align*}
    \underset{\theta}{\mathrm{min}} \: &\frac{1}{N} \sum_{\substack{\alpha,\sigma,r,V,\beta,\Bar{F}_y,\\\Bar{F}_x,\Bar{\mu F_z}  \in \mathcal{D}} }  \Big(F^\theta_{\mathrm{tot}}(\kappa, \mathrm{feat})- \Bar{F}_{\mathrm{tot}} \Big)^2 + (F^\theta_y - \Bar{F}_y)^2 \nonumber \\ 
    & + (F^\theta_x - \Bar{F_x})^2 +
    \lambda (\Bar{\mu F_z}- F_{\mathrm{tot}}^\theta(\kappa^\theta_+, \mathrm{feat}))^2
\end{align*}

\begin{remark}
    Firstly, by incorporating selected subsets of the measured states, $\mathrm{feat}$, in addition to the slip values, the proposed models are able to capture the intricate coupling between the effective lumped tire force curves and vehicle motion. While we made one choice for $\mathrm{feat}$, other selections are likely suitable, depending on the vehicle. Secondly, for fixed parameters $\theta$, \texttt{ExpTanh} requires only two evaluations of the function $\mathrm{exp}$, which is computationally cheap compared to the Magic Formula requiring three evaluations of $\arctan$; the gradient is also easier to compute.
\end{remark}
\section{Experiments}
We demonstrate the data efficiency, prediction accuracy, and computational efficiency of our tire models through several experiments: Comparisons to the Magic Formula and Fiala brush on the testbed vehicle dataset, and autonomous drifting on slalom and figure-8 trajectories. 

The experiments in this section were performed on the Toyota Supra \rev{described in~\cite{gohAVEC2022,balachandran2023human} and} heavily modified for high-performance autonomous driving. Vehicle state measurements are obtained from a commercial RTK GPS-INS unit at a rate of 250Hz. As a rear-wheel drive vehicle, the front tires operate in the pure slip regime with $F_{xf} = 0$, and the rear tires operate in the combined slip regime. 
\rev{We assume standard units for all quantities when they are not specified.}\looseness=-1

\addtocounter{figure}{1}
\begin{figure*}[b]
    \vspace*{-4mm}
    \centering
    \includegraphics{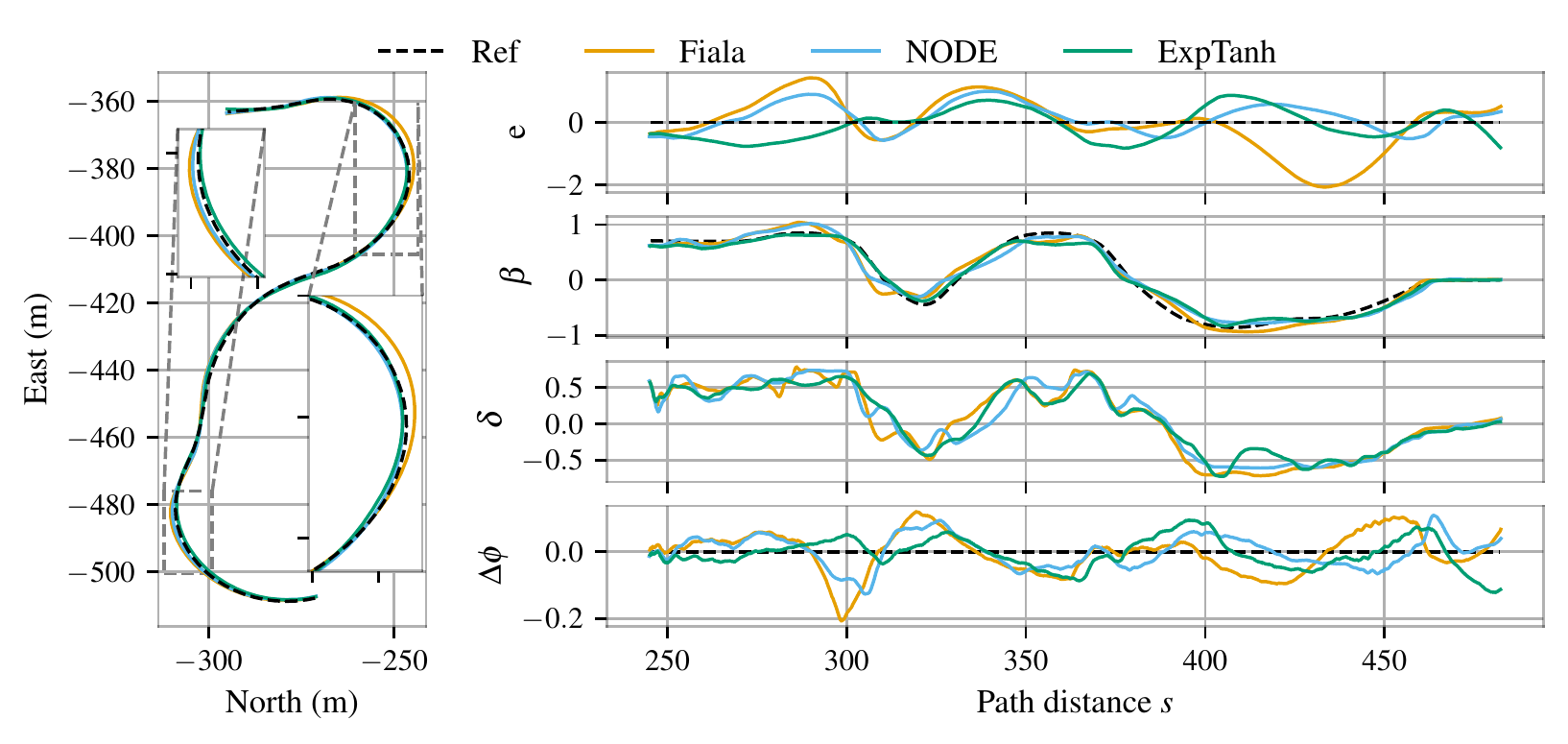}
    \vspace*{-4mm}
    \caption{Drifting on a slalom figure. Our approaches show better accuracy at trajectory tracking and \rev{fewer} steering oscillations than Fiala model.}
    \vspace*{-2mm}
    \label{fig:slalom-experiment}
\end{figure*}

\subsection{Evaluation of the Learned Tire Models}\label{subsec:learned-models}
To compare the tire models, we used a dataset $\mathcal{D}$ (Figure~\ref{fig:compare-learned-forces}) of manual and autonomous driving/drifting. The dataset contains $306887$ state measurements at 100 Hz, totaling $\sim1$ hour accumulated over the span of three months on the same surfaces \rev{under similar summer weather conditions.}

For the NODE model, $\mathrm{NN}^\theta_1$, $\mathrm{NN}^\theta_2$, and $\mathrm{NN}^\theta_{4}$ have $2$ hidden layers with $16$ nodes per layer, while $\mathrm{NN^\theta_3}$ have $4$ nodes per layer. For the $\texttt{ExpTanh}$ model, $\mathrm{NN}^\theta$ and $\mathrm{NN}^\theta_4$ have $2$ hidden layers with $3$ nodes per layer. All neurons used $\tanh$ as activation function. We used $\lambda = 0.01$ to express low confidence in the estimated $\Bar{\mu F_z} = 7000$. We trained the models via Adam optimizer \cite{kingma2014adam}, where the learning rate is set to decay exponentially with a rate of $0.01$ and an initial value of $0.001$. 
\rev{On a laptop with GeForce RTX 2060,} training both the pure slip $F_{yf}$ and the combined slip $F_{xr}$ and $F_{yr}$ models took $\sim27$ minutes for NODE, and only $\sim4$ minutes for $\texttt{ExpTanh}$.\looseness=-1

We compare our models with the Magic Formula and Fiala model. The parameters of the Magic Formula (Chapter 4, Section $4.3.2$ of \cite{PACEJKA2012xiii}) were obtained by optimizing a mean-square-error loss over the dataset. The Fiala model parameterization was empirically tuned by usage in the existing autonomous drifting NMPC framework ~\cite{balachandran2023human,Goh2019TowardAV}. 
Figure~\ref{fig:compare-learned-forces} summarizes our findings: Our tire models significantly improve prediction accuracy over the Magic Formula and Fiala while satisfying the tire fundamentals. The NODE model provides the best prediction accuracy while taking significantly more time to train and evaluate. In contrast, $\texttt{ExpTanh}$ achieves slightly lower prediction accuracy compared to the NODE model while being easy to train and evaluate.\looseness=-1

Figure~\ref{fig:exptanh-forces} shows how the learned $\texttt{ExpTanh}$ model efficiently captures the coupling with the vehicle states $r, V, \beta$. This suggests that our models not only fit the tires but can also incorporate complex chassis interactions (e.g., weight transfer and suspension dynamics). In the pure slip model, $r$ shifts the center of the curve. This could be due to significant static and dynamic camber from the test vehicle's aggressive drift-specific front suspension setup. Low speed values tend to flatten the curve, while the slip angle corresponding to the peak force decreases with increasing sideslip angle $\beta$. For the combined slip model, the dependency on $r,V$ is most significant in the nonlinear transitional region at low longitudinal slip. As expected from the tire fundamentals, the magnitude of $F_{xr}$ decreases as the slip angle increases for fixed $\sigma_r$.
\addtocounter{figure}{-2}
\begin{figure}[!hbt]
    \centering
    \vspace*{-3mm}
    \includegraphics{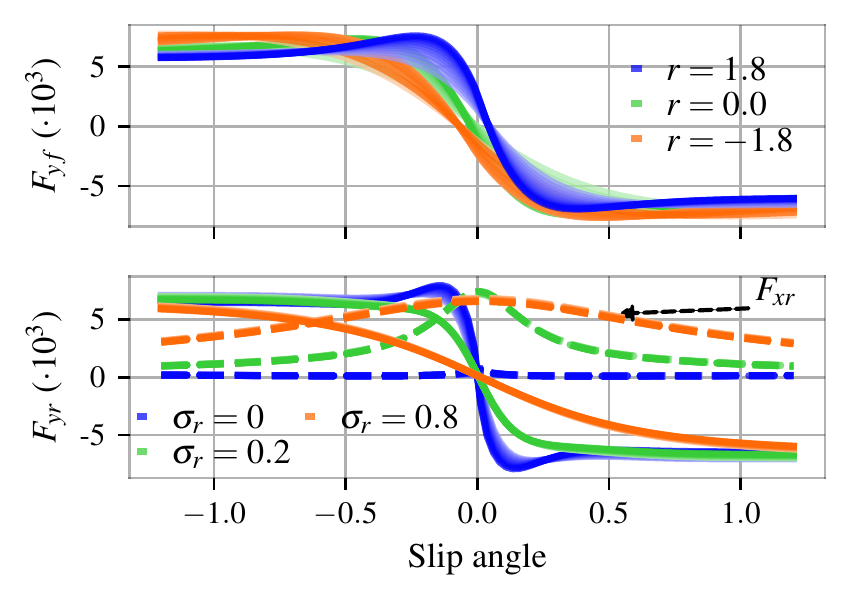}
    \vspace*{-8mm}
    \caption{Impact of the states $r,V,\beta$ on the learned \texttt{ExpTanh} model. For the front tire, the blue curve corresponds to fixed $r=1.8, \beta=0.9$, and $V$ ranging from $5$ to $20$ with lower values represented by lighter colors. The green curve follows the blue curve but with $r=0$. The orange curve uses fixed $r=-1.8, V=12$, and $\beta$ varying from $-0.9$ to $0.9$. For the rear tire, $V$ ranges from $5$ to $20$ with fixed $r=-1.8$ for the blue curve, $r$ ranges from $-1.8$ to $1.8$ with fixed $V=12$ for the green curve, and both $V$ and $r$ vary for the orange curve. }
    \vspace*{-7mm}
    \label{fig:exptanh-forces}
\end{figure}

\subsection{Autonomous Drifting with Learned Tire Models}
To evaluate their practical closed-loop performance, we use our \rev{learned models in Figure~\ref{fig:compare-learned-forces}} as direct drop-in replacements for a Fiala model in an existing closed-loop NMPC framework for autonomous drifting ~\cite{gohAVEC2022,balachandran2023human}. The reference trajectories were pre-computed via nonlinear optimization with the benchmark Fiala model. The NMPC cost function primarily penalizes the lateral error $e$, the deviation from the reference sideslip angle $e_\beta = \beta - \beta_{ref}$, and the relative deviation $\Delta \phi$. \rev{The sideslip $\beta_{ref}$ enforces the drifting profile.}

\subsubsection{Drifting on Slalom Trajectory}
For the first experiment, we compare the closed-loop performance of the benchmark Fiala, NODE, and $\texttt{ExpTanh}$ models on a transient slalom trajectory (Figure~\ref{fig:slalom-experiment}). The integrated NODE formulation was approximated with a neural network trained on its output. The slalom trajectory has corners with reference sideslip angle of up to $43^\circ$ and velocity between \rev{$31$mph} and \rev{$45$mph}. Figure~\ref{fig:slalom-experiment} demonstrates improved tracking performance: In terms of root mean squared error, $\texttt{ExpTanh}$ tracks the path ($e$ and $\Delta \phi$) up to $3.5\times$ better than the Fiala model while achieving up to $1.5\times$ better sideslip tracking performance. The NODE model achieves slightly lower performance than $\texttt{ExpTanh}$, possibly due to the loss of accuracy from the approximation procedure. Importantly, we also note fewer steering oscillations $\delta$ when $\texttt{ExpTanh}$ and NODE are used, as compared to the baseline Fiala model. 

\subsubsection{Autonomous Drifting with 3 Minutes of Data}
\addtocounter{figure}{1}
\begin{figure*}[t]
    \includegraphics{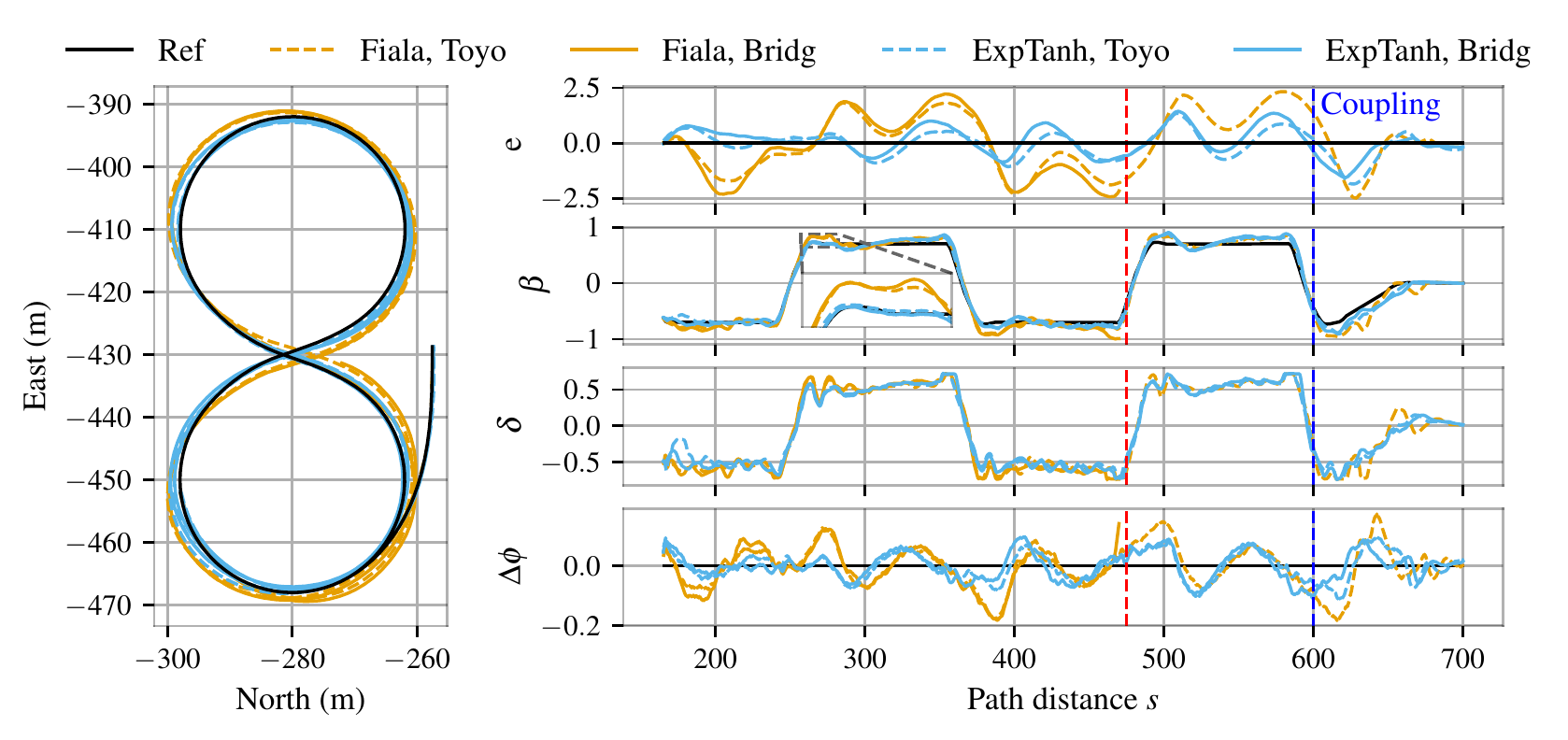}
    \vspace*{-4mm}
    \caption{Drifting on a \rev{figure-8 trajectory} with $3$ minutes of data. \texttt{ExpTanh} shows better tracking performance with both tires, especially in transitional regions. The red line indicates where the \emph{Bridgestone} + Fiala test was ended due to the safety driver feeling uncomfortable with the tracking error. }
    \vspace*{-4mm}
    \label{fig:figure-8-experiment}
\end{figure*}
\newcommand{\exptanh}{$\texttt{ExpTanh}$ }
This set of experiments investigates the generalizability of the ExpTanh model. First, we perform experiments on a \rev{figure-8 trajectory} with both the benchmark Fiala model and the $\texttt{ExpTanh}$ model. We then changed the rear tires from Toyo Proxes Sport 275/35R18, which we used for all previous tests, to Bridgestone Potenza Sport 275/35R18. A safety driver then manually drove the car on the skidpad, with unstructured grip and drift maneuvers, for $\sim 3$ minutes. This data was then used to train an $\texttt{ExpTanh}$ model; this took $<15s$ on a laptop with a GeForce RTX 2060. This freshly fitted model was then again compared to the benchmark model on the same \rev{figure-8 trajectory}. 

Figure~\ref{fig:figure-8-experiment} summarizes our findings. For the original \emph{Toyo} tires, performance was significantly better with the \exptanh model than the baseline Fiala model. Importantly, the closed-loop behavior with the \exptanh model was similar after switching to the new \emph{Bridgestone} tires and retraining the network. In contrast, the performance with the unchanged Fiala model significantly degraded, showing that there was indeed a notable difference in the behavior of the tires that the \exptanh model successfully adapted to with sparse data. This is also reflected in the root mean squared $e$ and $e_\beta$ values: compared to the baseline, we see a $>4\times$ improvement in lateral error, and $>2\times$ improvement in sideslip tracking for both tires.
\begin{center}
\begin{tabular}{ |c|c|c|c|c|  } 
  \hline
  & \multicolumn{2}{|c}{Toyo} & \multicolumn{2}{|c|}{Bridgestone}\\
 RMS & Fiala & $\texttt{ExpTanh}$ & Fiala & $\texttt{ExpTanh}$\\ 
  \hline
 $e$ (m) & $1.77$ & $0.40$ & $2.02$ & $0.27$\\ 
 $e_{\beta}$ (rad) & $0.013$ & $0.006$ & $0.011$ & $0.004$\\ 
 \hline
\end{tabular}
\end{center}

With $\texttt{ExpTanh}$, the controller tracks the sideslip $\beta$ reference with less overshoot and less steering oscillation compared to the Fiala model. 
This difference is particularly pronounced at the end of each transition, as shown in the zoomed section on the sideslip evolution, where we expect to see more complex interactions between the vehicle states due to transient load transfer and high yaw rates. This suggests not only that the  $\texttt{ExpTanh}$ model is able to capture these effects but also that the controller benefits by exploiting this in closed loop. In contrast, the controller with the Fiala model tends to overshoot severely during these transitions. 

Another region with complex coupling is the slow transition from drifting to grip driving at the end of the experiment  (e.g. $s \in [600, 660]$). Here, the baseline \emph{Toyo} + Fiala model combination exhibits steering and sideslip oscillations. In contrast, with both tires, $\texttt{ExpTanh}$ smoothly tracks sideslip, and has better lateral error performance.

\begin{figure}[!hbt]
    \centering
    \vspace*{-4mm}
    \includegraphics{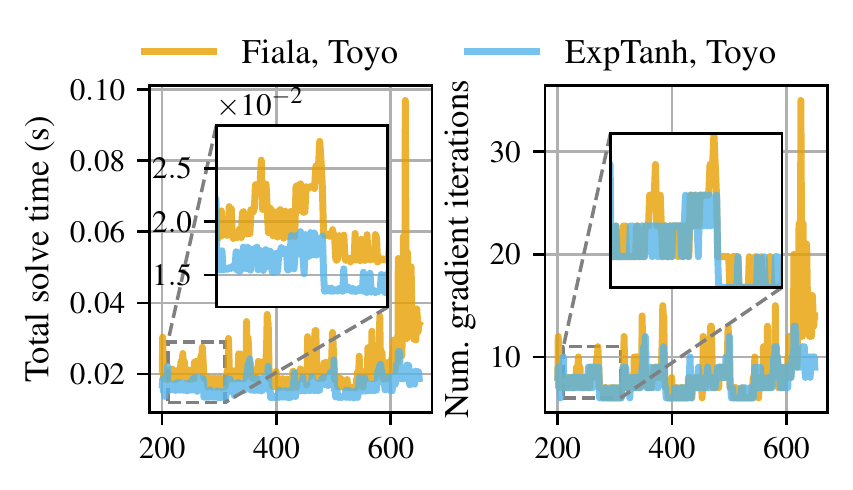}
    \vspace*{-8mm}
    \caption{Compute time and number of gradient iterations from the controller.}
    \vspace*{-2mm}
    \label{fig:exptanh-cpu}
\end{figure}

Figure~\ref{fig:exptanh-cpu} compares the observed  optimal control problem computation time during the \emph{Toyo} experiments. Due to its simplicity, forward evaluation of the Fiala model is likely faster than \texttt{ExpTanh}. However, in NMPC, the fidelity of the tire model and the smoothness of its Jacobian are important. Figure~\ref{fig:exptanh-cpu} shows that NMPC with Fiala often needs more gradient iterations than \texttt{ExpTanh} to converge to a solution. This is exacerbated in the transitional regime, $s\geq 600$, where the number of iterations triples with the Fiala model, but remains similar with \exptanh. This suggests that NMPC with \texttt{ExpTanh} can be both faster and more consistent. 

\section{Conclusion}\label{sec:conclusion}
We propose a family of tire force models based on neural ordinary differential equations (NODE) and $\texttt{ExpTanh}$. These models combine physics-based tire modelling fundamentals with the ability to directly learn, using onboard sensor measurements, higher-order effects from the interaction between the tires, suspension, and road. Autonomous drifting experiments, which subject the model to extreme conditions, demonstrate improved tracking performance, optimized computation time, and unprecedented data efficiency: Learning with only $3$ minutes of driving. Finally, our rapid training time (usually a few seconds) suggests that future work could explore using these models in a life-long learning setting, where the tire curves are updated online during driving.

\textbf{Acknowledgements.} The authors would like to thank our colleagues at Toyota and UT Austin's Autonomous Systems Group for their generous support. We thank ONR, AFOSR, and NSF for their support on our prior works on NODE.\looseness=-1

\clearpage
\newpage


\bibliographystyle{ieeetr}
\bibliography{bibliography}
\end{document}